# Droplet evaporation residue indicating SARS-COV-2 survivability on surfaces

Zilong He [1,2], Siyao Shao[1,2], Jiaqi Li[1,2], S. Santosh Kumar [1,2], J. B. Sokoloff [3,4] and Jiarong Hong[1,2*]

1. Department of Mechanical Engineering, University of Minnesota, Minneapolis, MN 55455
2. Saint Anthony Falls Laboratory, University of Minnesota, Minneapolis, MN 55414
3. Department of Physics, Northeastern University, Boston, MA 02115
4. Department of Physics, Florida Atlantic University, Boca Raton, FL 33431

*Corresponding author: jhong@umn.edu

**Abstract:** We conducted a systematic investigation of droplet evaporation on different surfaces. We found that droplets formed even with distilled water do not disappear with evaporation, but instead shrink to a residue of a few micrometers lasting over 24 hours. The residue formation process differs across surfaces and humidity levels. Specifically, under 40% relative humidity, 80% of droplets form residues on plastic, uncoated and coated glass, while less than 20% form on stainless steel and none on copper. The formation of residues and their variability is explained by modeling the evaporation process considering the presence of nonvolatile solutes on substrates and substrate thermal conductivity. Such variability is consistent with the survivability of SARS-CoV-2 measured on these surfaces. We hypothesize that these long-lasting microscale residues can potentially insulate the virus against environmental changes, allowing them to survive and remain infectious for extended durations.

The ongoing COVID-19 pandemic has infected more than thirty million people as of now, causing major disruption to the global economy and social order. It has been well accepted that the virus causing the disease, severe acute respiratory syndrome coronavirus-2 (SARS-CoV-2), can be transmitted through contact of virus-laden respiratory droplets on surfaces. Particularly, studies have found much higher concentration of SARS-CoV-2 RNA deposited on surfaces in hospitals rather than as aerosols [1,2], pointing to the importance of investigating the virus survivability on surfaces. As reported by two recent experiments [3,4], SARS-CoV-2 has a long survival time on different surfaces and can remain viable under different temperature and humidity levels. Specifically, Chin *et al.* [3] investigated the stability of SARS-CoV-2 deposited as droplets on ten surfaces at 60% relative humidity (RH) with variation in temperatures, and found the virus to be more stable on smooth surfaces (e.g. glass and plastic), remaining viable for up to two to four days, respectively with survival time decreasing at higher temperatures. Similarly, Van Doremalen *et al.* [4] found virus survival time on four surfaces, at 40% RH, to vary from approximately seven hours on copper to more than three days on plastic (polypropylene). However, no study so far has provided any physical mechanisms that can explain the long survival times, the large variation between the different surface materials tested, as well as the impact of environmental changes on surface transmission. Such mechanisms, related to droplet evaporation process, can be critical for understanding the carriage and transmission of SARS-CoV-2 as summarized in a recent review paper [5]. Here we hypothesize the evaporation characteristics of respiratory droplets may indicate SARS-CoV-2 survivability on different surfaces and under different humidity and temperature conditions. In the literature, the studies of droplet evaporation on surfaces typically involve seeded particles and focus on particle pattern formation for various applications such as inkjet/3D printing, manufacturing self-assembled structures, etc. [6]. Only one study investigated the evaporation of ultrapure water droplets on hydrophobic substrates that generates submicron residues [7]. There is no systematic study of such water droplet evaporation on different surfaces of interest, nor works that make connection between virus transmission and droplet evaporation.

In this study, we conduct a systematic experiment to assess the evaporation process of distilled water droplets on surfaces with deposited droplet size ranging from 5 to 100 μm, within the range of respiratory droplets generated by human breathing and speaking [8]. The distilled water is selected instead of respiratory droplets to minimize the variability of droplet chemical content on our test results. Additionally, test surfaces are chosen to match those used in [3] and [4]. A detailed description of the experiment is provided in the supplementary materials.

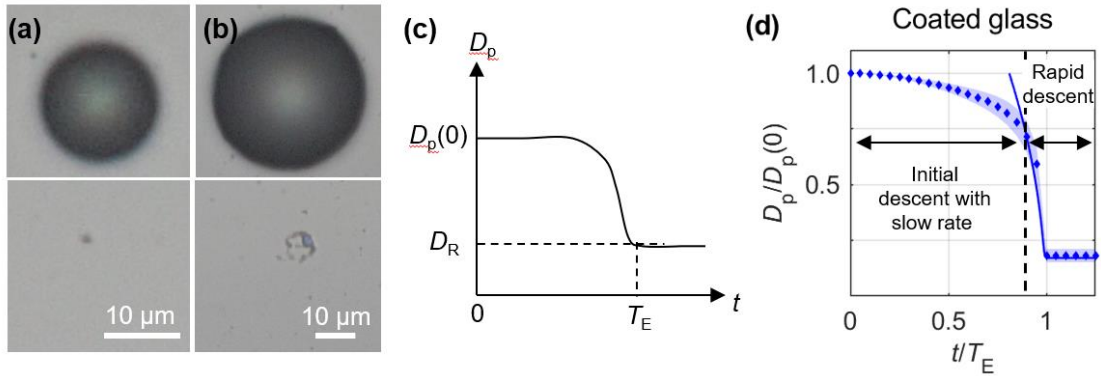

**FIG. 1.** Initial and final frames from videos of droplets producing (a) single residue and (b) multiple residues. (c) Schematic of the evaporation curve illustrating the variation of droplet size with time. $D_p(0)$ is the initial droplet diameter, $T_E$ the time at which the droplet forms the residue of size $D_R$. (d) The normalized evaporation curve calculated by averaging 100 individual droplets evaporating on the coated glass surface at a temperature of 22 ºC and humidity of 40% RH. The measured time varying size from the images are used as sample points to generate a continuous evaporation curve at discrete time steps through piecewise Hermite polynomial interpolation. The standard deviation indicating the differences between the sampled droplets is presented as the shading around each data point and the evaporation model by the solid line.

We found that during evaporation, droplets on the tested surfaces first shrink in height (Constant contact radius mode) and then in diameter (constant contact angle mode) to form a thin liquid film, usually leaving behind a single residue on the order of micrometers (Fig. 1a and Video S1). Sometimes, the film can break up into multiple residues (Fig. 1b and Video S2) which can persist for hours with no visible change in their size. Such residues appear in different forms (Fig. S1 and Video S3-S6) on all surfaces except copper, at 40% RH. On the copper surface, only a faint signature of a residue can be seen, suggesting a film with a thickness below our resolution limit (~300 nm), much smaller than those for other surfaces. To quantify the droplet evaporation process, we measure the diameter ($D_p$) as a function of time ($t$) for the different surfaces (Fig. 1c). We define $D_p$ as the area equivalent diameter of the droplet to enable comparisons between non-spherical and spherical shapes observed. The initial droplet size $D_p(0)$ is measured at the start of evaporation when the droplet begins to change in size or height. The evaporation time $T_E$ is defined as the time at which the droplet shrinks to residue size $D_R$, i.e., $D_p(T_E)=D_R$. In cases where the droplet disappears completely we set $D_P(T_E)=0$, while for cases with multiple residues, we measure $D_R$ is defined as the root mean square of the individual residue sizes. To characterize the general evaporation trend of droplets of different sizes, the evaporation curves are normalized using the $D_P(0)$ and $T_E$ corresponding to each droplet. The initial droplet diameter $D_P(0)$ and evaporation time $T_E$ yield approximately a linear relationship under our experimental conditions for all surfaces except copper for which $T_E$ shows little dependence on $D_P(0)$ (Fig. S2). For the coated glass surface (Fig. 1d), the evaporation curve exhibits an initial slow rate of change in size over a duration of ~$0.8T_E$ followed by a rapid descent to form the final residue, of about 18% of $D_P(0)$.

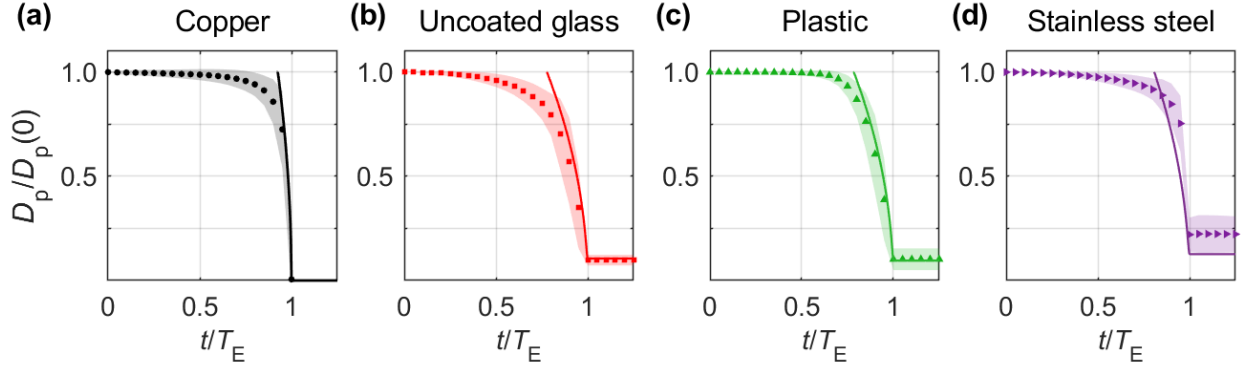

**FIG. 2.** Normalized evaporation curves on (a) copper, (b) uncoated glass, (c) plastic, and (d) stainless steel surfaces at a temperature of 22 °C and humidity of 40% RH. The curves are obtained following the same procedure as that for Fig. 1d. Solid line: Simulation results

Compared with the coated glass surface (Fig. 1d), the evaporation curves for the other surfaces show a similar trend in general (Fig. 2). However, the evaporation rate and residue size vary among different surfaces, depending on the surface properties including wettability, roughness and thermal conductivity. Specifically, coated glass that has strong hydrophobicity and smoothness presents the highest initial evaporation rate. The metal surfaces (i.e., copper, and stainless steel) with higher thermal conductivity exhibit a steeper change in size near the end of evaporation, compared to plastic and both glass surfaces with low thermal conductivity. The copper substrate does not yield any resolvable residue at 40% RH, while the residues for the other surfaces fall within in the range of 9-22% of $D_P(0)$. The rougher surfaces like plastic and stainless steel show larger variation in residue size compared to the smoother glass surfaces.

The formation of microscale residues from pure water evaporation was reported in 9, which suggests that this phenomenon is a result of deliquescence by ionic compounds in the photoresist substrate. However, such a mechanism cannot explain the observations from the current experiment using substrates without similar ionic compounds. Here we attribute the formation of residues to the presence of nonvolatile solutes on substrates which gradually dissolve into the droplet near the contact line during the evaporation. The dissolution of such nonvolatile content slows down and eventually ceases evaporation, leaving residues on substrates.

A physical model of this evaporation process (Eq. 1) is proposed by including effects of both non-volatile solute [9], and substrate conductivity [10] on the quasi steady evaporation rate equation proposed by Hu & Larson [11].

$$\dot{m} = \frac{\rho dV}{dt} = -\frac{\pi D_p D}{2}\left(1 - \frac{\phi(t)D_0^3}{D_p(t)^3} - \text{RH}\right)C_s\left(0.27\theta^2 + 1.3\right) \times M \qquad (1)$$

where, $\rho$ is the density, $D_p$ the wetted diameter of the droplet, $V$ the volume, $D$ the diffusion coefficient of water vapor, $\phi$ the volume fraction of the solute (evaluated by Nernst and Brunner equation), $D_0$ the initial droplet wetted diameter, RH the relative humidity, $C_s$ the saturation vapor concentration at the liquid-gas interface, $\theta$ the contact angle, and $M$ a correction factor between 0 and 1, coupling substrate conductivity and evaporative cooling of the droplet [10]. The dissolution of the nonvolatile solute is described by the Nernst and Brunner equation [12]

$$\frac{dC(t)}{dt} = \frac{D_s A(t)}{V(t)h_d}(C_{sn} - C(t)) \qquad (2)$$

where $C$ is the concentration of solute inside the droplet, $D_s$ is the diffusion coefficient of the solute in the solvent, $h_d$ is the thickness of the diffusion layer. $A$ is the area near contact line, $V$ is the volume of the droplet, and $C_{sn}$ the solubility of the solute.

Note that since the diameter $D_p$ is much smaller than the capillary length (2.37 mm), the effect of gravity is neglected, resulting in the droplet shape resembling a spherical cap with volume given by $V = \pi h(3D_p^2 + D_p \tan(\theta/2)^2)/24$. The values for $D$ and $C_s$ are evaluated by equations from Kumar et al [13]. while the contact angle $\theta$ is taken from prior studies with similar experimental conditions [14–16]. Esmaili et al. experimentally shows that when droplets impact on the substrates with particles, the particles migrates naturally to the contact line, supporting our hypothesis that the dissolution happens near the contact line [17]. The concentration of the solute calculated is converted to is the volume fraction $\phi$ using $\phi = C/\rho_{solute}/(C/\rho_{solute} + V)$. The properties of the solute ($C_s$, $\rho_{solute}$, $D_s$, $h_d$) and $M$ are treated as fitting parameters. Netz derived an analytic expression for the residue size as a function of the solute volume fraction and the relative humidity $D_R = D_p(0)(\phi_0/(1-\text{RH}))^{1/3}$ where $\phi_0$ is the initial solute concentration. [9]

We compare the time scale of dissolution ($\tau_{dis} \sim dD_p/D_s$) and evaporation time found from [9,10]

$$\frac{dD_p}{dt} = M \times \frac{24D}{\rho \tan\frac{\theta}{2}\left(3 + \tan^2\frac{\theta}{2}\right)D_p}\left(1 - \frac{\phi(t)D_0^3}{D_p(t)^3} - \text{RH}\right)C_s\left(0.27\theta^2 + 1.3\right) \qquad (3)$$

The volume fraction of solute is allowed to increase only in the limit where dissolution is much faster than evaporation ($dt \gg \tau_{dis}$). The coupled equations (Eq.2 and 3) are solved numerically in MATLAB with a constant contact angle assumption except for coated glass which we assume a linear decreasing contact angle starting from ~$0.8T_F$ [18] the results for which are shown in Fig.1(d) and Fig. 2. The time of constant contact radius mode is obtained from our experiments. The slow evaporation process on low conductivity substrates and high RH environment allows more solute to dissolve, leading to larger residues.

The maximum relative error is 15% for coated glass, 11% for copper, 15% for uncoated glass, 7% for plastic and 40% for stainless steel. The large relative error on stainless steel can be attributed to small residue fraction and the rough surfaces that makes it hard to evaluate the nonspecial droplet equivalent diameter and residual sizes. Our model has 2 limitations. It fails to predict the fraction and the eventual decay of residues on different substrates. In addition, the contact angle starts to slowly decrease as the droplet shrinks to a residue, making our model deviate from experiments.

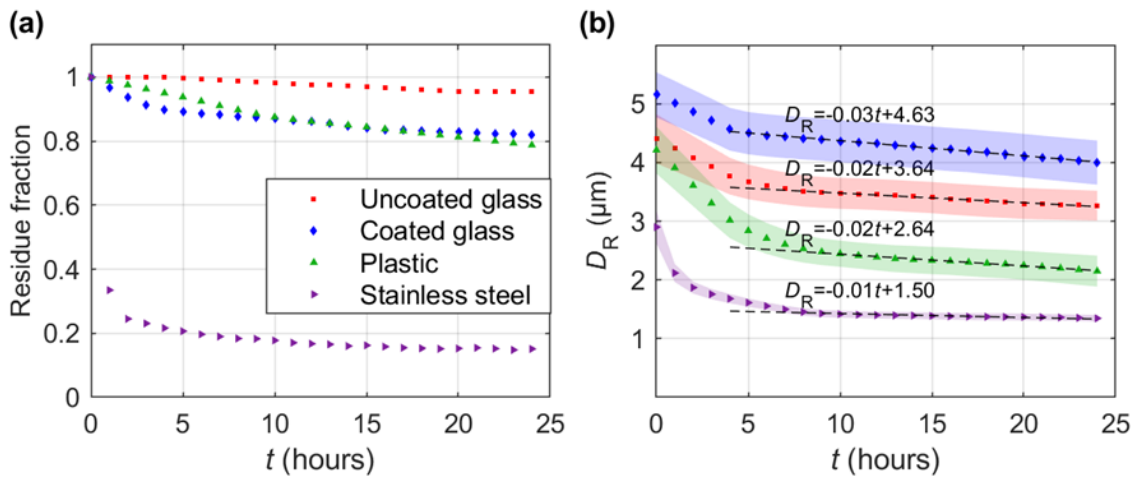

**FIG. 3.** Long term stability of residues on various test surfaces measured at a temperature of 22 °C and humidity of 40% RH. (a) Residue fraction as a function of time on each substrate. (b) Average area equivalent diameter $D_R$ of residues sampled over the same duration with shaded region representing the standard error, and the dashed lines indicate linear least square fit conducted over a range of $t$ near the end of each data set where a linear trend can be clearly observed, from above ~5 hours for coated glass to data above ~8 hours for the remaining.

The resolvable residues exhibit a stability in number and size for a period of 24 hours as shown in Fig. 3. Specifically, the percentage of residues that remain, referred to as residue fraction, decays gradually with time for all surfaces except for stainless steel which displays a sharp decline at the beginning, reaching a plateau at ~15% potentially due to the relatively higher thermal conductivity and a larger contact area associated with surface roughness. The uncoated glass retains the highest residue fraction (~95%), while the coated glass and plastic both yield a lower fraction of ~80% after 24 hours. The drop in residue fraction can be attributed to the evaporation of smaller residues present on these surfaces as indicated by the larger variability in residue size seen in Fig. 2. The average residue size (Fig. 3b) for all surfaces show a relatively larger decrease within the first few hours, followed by an almost linear decay with a very shallow slope (-0.01 to -0.03 μm /hour) at longer durations, indicating their survival time could extend well beyond 24 hours.

Once formed, these residues show strong durability even under fluctuations of ambient temperature and humidity. They can stay on plastic and glass surfaces even after the surfaces are treated with a heat gun for 60 s at a temperature of ~60 °C (measured at the surface), while the same treatment removes more than ~90% of residues on stainless steel, possibly due to its higher thermal conductivity. In comparison, we found that wiping is more effective for residue removal across all surfaces (applying Kimtech wipes for 10 s can remove >95% of the residues).

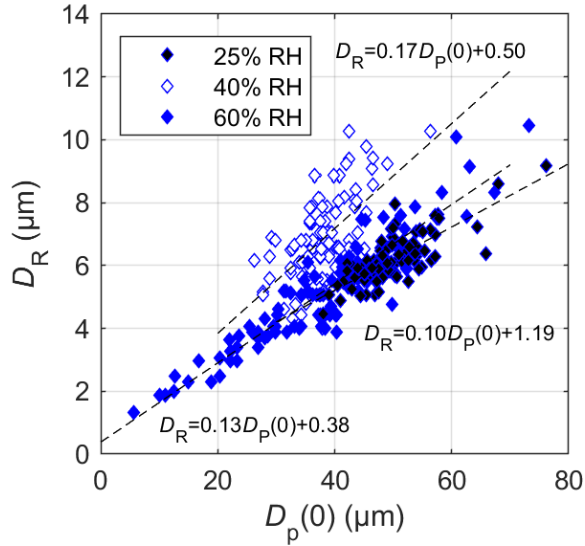

**FIG. 4.** Variation of residue size $D_R$ with initial droplet size $D_P(0)$ at 22 °C and three humidity levels (25%, 40% and 60% RH) on a coated glass substrate. Lines indicate linear least squared fits to the data.

We found that the residue formation process is strongly influenced by the ambient humidity. As the humidity increases from 25% RH to 60% RH, the fraction of droplets that form residues increases by ~5% on the coated glass surface, ~15% on the plastic surface with no significant change observed on the plastic and stainless steel surfaces. More importantly, at 60% RH we also observe the formation of residues on the copper surface, although for a much lower fraction of droplets. For the coated glass substrate (Fig. 4), the residue size scales linearly with the initial droplet size at all humidity values with very similar slopes. Specifically, the minimum droplet size that can form a residue decreases with humidity, from ~30 μm at 25% RH to ~5 μm at 60% RH. We observe similar trends between the two humidity values for the other surfaces (Fig. S3).

Overall, our findings provide a physical mechanism contributing to the long survival time and stability of viruses under practical settings. Specifically, we suggest that the residues with size 1-2 orders larger than that of SARS-CoV-2 found in our experiments can serve as a shield, insulating the virus against extreme environmental changes [8]. Furthermore, the presence of a lipid bilayer with a hydrophilic outer surface on the virus [19], allows them to remain stable in high humidity found within residues. Accordingly, the probability of forming residues and their stability can indicate the virus survivability on different surfaces. For instance, the residues are found to be much more difficult to form on copper, which shows the shortest survival time of SARS-CoV-2 in [4]. Compared with plastic, the stainless steel has lower probability of sustaining the formed residue for long term at 40% RH, mirroring the survivability results for plastic and stainless steel reported in [4].

The physical insights gained from our work can be extended to other viruses that are transmitted through respiratory droplets (e.g., SARS/MERS viruses, flu viruses, etc.), particularly, to SARS-CoV-1 which has a survivability trend very similar to those of SARS-CoV-2 on different surfaces [4]. Our findings suggest that high temperature (through enhancing evaporation rate) and low humidity can inhibit the formation of residues, lowering the survivability of viruses on surfaces. Regarding temperature effects, such inference is consistent with reduced survivability of virus with increasing temperature reported in multiple studies [3,20,21]. However, despite a number of studies investigating the humidity effect on virus survivability on surfaces [20,21], their experiments were conducted using virus-laden droplets of ~mm size, which forms residues at all humidity conditions tested according to our study. Therefore, the probability of residue formation cannot be used to explain the variation of virus survivability with humidity in their studies, which are likely caused by other mechanisms. The adverse effect of humidity on virus infectivity reported in the literature [22,23] points largely to airborne transmission, which can be explained by increased aerosol settling at higher humidity through condensation, and is not relevant to the mechanism discussed in our study.

Our tests show that wiping with regular water-absorbent tissue paper can remove more than 95% of the residues on surfaces if disinfecting wipes are not available. Particularly, our results derived from the experiments using droplets with size matching those generated during human breathing and speaking has specific implications for COVID-19, which displays an exceedingly high rate of spread than earlier viruses, associated with high viral loads in the upper respiratory tract and potential transmission by asymptomatic/presymptomatic individuals [24–26]. Our results suggest that even tiny droplets (<20 µm) can leave residues under moderately high humidity (>40%), causing significant spread of virus through surface contamination. Therefore, our study highlights the importance in wearing masks under such conditions towards minimizing the spread of virus to surfaces through normal respiratory activities e.g., breathing and speaking [27]. In addition, lowering the indoor humidity when possible can suppress the formation of such residues (e.g., significant drop in fraction of residue forming droplets in steel below 15% RH and below 10% RH for other surfaces), and limit the spread of viral infection through contact from such small respiratory droplets, as we continue to reopen our economy and workplaces in the future.

In the end, we would also like to caution the readers from generalizing the quantitative results (e.g., evaporation rate, residue fraction, etc.) present in our experiments, since they are dependent on specific surface and environmental conditions. Accordingly, it would be of practical significance to investigate the evaporation residues over a broader range of surface substrates and under different environmental factors (e.g., humidity, temperature, etc.), which can lead to actionable prevention measures to reduce the virus transmission through contaminated surfaces. Our work can potentially inspire a host of future research using more advanced diagnostic, analytical and simulation tools to elucidate the formation and characteristics of residues and their connection with virus transmission.

We acknowledge the support from the University of Minnesota for this research. We would also like to thank Dr. David Pui for the equipment support, Dr. Suo Yang and Dr. Lei Feng for fruitful discussion of the results and Barbara Heitkamp for help editing the manuscript.

**Droplet evaporation residue indicating SARS-COV-2 survivability on surfaces**

Zilong He [1,2], Siyao Shao[1,2], Jiaqi Li[1,2], S. Santosh Kumar [1,2], J. B. Sokoloff [3,4] and Jiarong Hong[1,2*]

*1.Department of Mechanical Engineering, University of Minnesota, Minneapolis, MN 55455*
*2.Saint Anthony Falls Laboratory, University of Minnesota, Minneapolis, MN 55414*
*3. Department of Physics, Northeastern University, Boston, MA 02115*
*4. Department of Physics, Florida Atlantic University, Boca Raton, FL 33431*

*Corresponding author: jhong@umn.edu

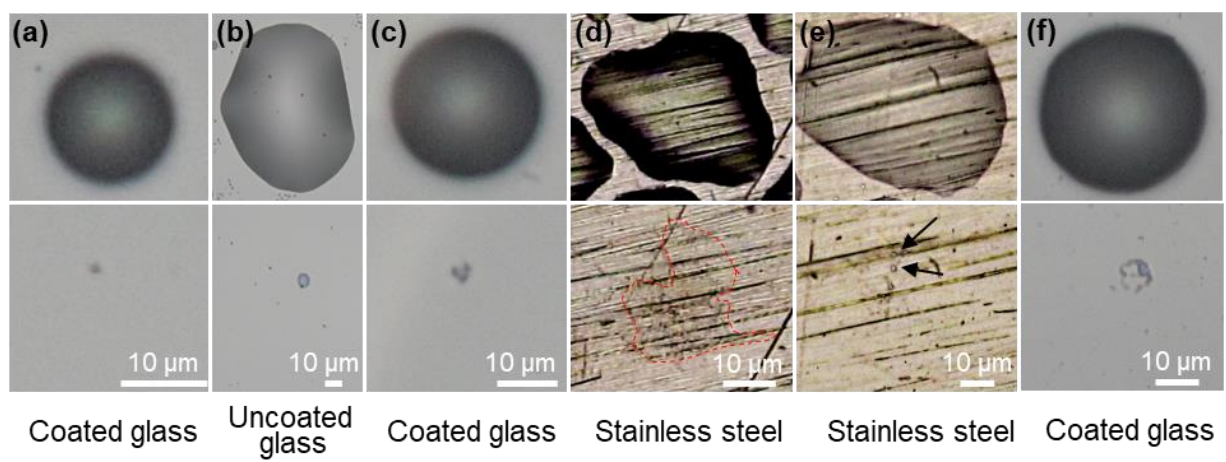

**FIG. S1.** A gallery of original droplets (upper) and their corresponding residues (lower) indicating the various morphologies of residues formed. Single residues form by (a) non-pinning droplet evaporating on a coated glass surface, (b) pinned droplet evaporating on an uncoated glass surface, (c) film recoil for pinned droplet and (d) contact pinned evaporation on stainless steel forming a large area of residue (marked by outline). Multiple residues form due to (e) roughness induced film break-up in stainless steel surface (with arrows marking the individual residues) or (f) surface tension induced film breakup on a coated glass substrate.

The evaporating water droplets, at ambient conditions of ~22ºC and ~40% RH, do not disappear but leave a residue that persists for hours with no visible change under the different surfaces tested, except for copper. Figure S1 illustrates the different types of residues observed in our experiment. We either obtain a single residue, most likely a thin film or droplet, or multiple residues formed by breakup of a thin film. Single residues form through evaporation on a glass surface both in the absence of surface adhesion for a hydrophobic surface (Figure S1a) or with on a hydrophilic surface with strong adhesion (Figure S1b). Near the end of evaporation on a coated glass substrate, sometimes the thin liquid film recoils due to effect of surface tension, leaving behind a larger concentrated residue in the middle (Figure S1c). Alternatively, on a stainless steel surface, strong hydrophilic behavior of the evaporating droplet results in a large area thin film residue (Figure S1d). We do observe similar thin films on copper substrates but with a thickness much smaller than for stainless steel. Our approach is thus unable to fully quantify the residues size on copper surfaces due to the weaker signal inherent to such thin films at this humidity level. Finally, the formation of multiple residues is often through breakup of a pinned film due to surface roughness, e.g., on stainless steel (Figure S1e) or surface tension instabilities e.g., on coated glass (Figure S1f).

We repeated our experiments with droplets condensed from human breath instead of a nebulizer. The results also show the formation of similar stable residues persisting for a long term, with the same qualitative trends across the different surfaces. Such results point to the strong relevance of our experiment to disease transmission through respiratory exhalation, although they are not presented here in a quantitative fashion, considering the large variability of the chemical contents in human breath.

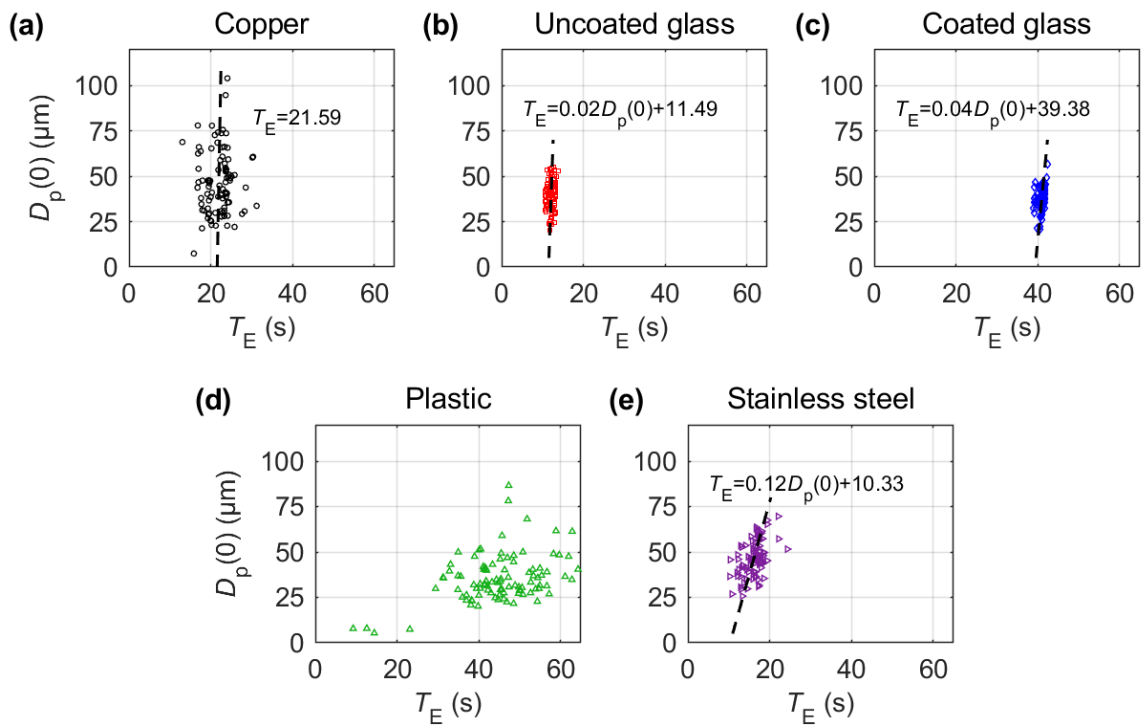

**FIG. S2.** Variation of droplet evaporation time ($T_E$) as a function of initial droplet size $D_p(0)$ for (a) copper (b) uncoated glass (c) coated glass (d) plastic and (e) stainless steel surfaces. Dashed line indicates least squares linear fit between $D_p(0)$ and $T_E$.

We observe an approximately linear trend between evaporation time and initial droplet size, but with a slope that varies strongly across the different surfaces, from ~0.12 for stainless steel to ~0.02 for the uncoated glass surface. Interestingly, our measurements on the copper surface shows no clear dependence between the droplet size and evaporation time, possibly due to the high thermal conductivity influencing the evaporation process. The plastic surface, on the other hand, does not show a clear trend in the measurements and also takes the longest time for evaporation, on average, followed by the coated glass. Such trends compare favorably to lower evaporation rates expected on hydrophobic surfaces due to the smaller surface area exhibited by the droplet. In contrast, all hydrophilic surfaces measure evaporation times which are approximately half of the hydrophobic glass, with the uncoated glass showing even faster evaporation. The large scatter in the data for copper, stainless steel and plastic cases can be attributed to the variation in droplet shapes and size as well as the variety of residue types formed on those surfaces, which points to the presence of multiple evaporation mechanisms. Surfaces with minimal variation in droplet residue type i.e., both glass surfaces, show the least amount scatter from the linear trend.

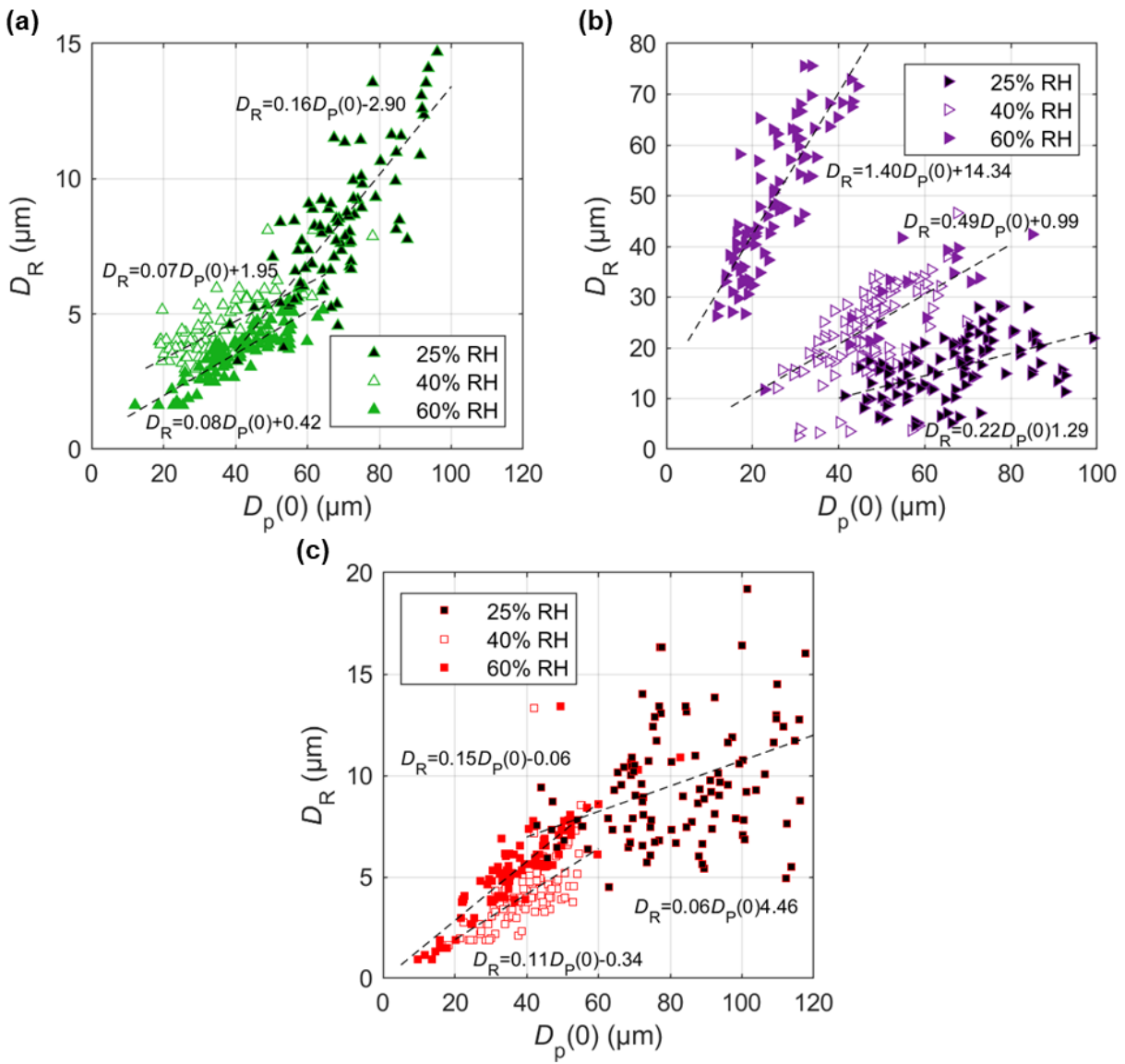

**FIG. S3.** Variation of residue size $D_R$ with initial droplet size $D_P(0)$ at 22°C and two humidity levels (25%, 40% and 60% RH) on (a) plastic, (b) stainless steel and (c) uncoated glass substrates. Lines indicate linear

least squared fits to the data. For the stainless steel surface at 60% RH, the smaller residue size clusters are neglected when estimating the trend line.

The increase in humidity from 25% RH to 60% RH leads to an increase in the fraction of residue forming droplets, with coated glass increasing from 55% to 90%, plastic from 5% to 30% and copper from 0% to 15% (i.e., no residues to residues at higher humidity). On the other hand, the stainless steel and uncoated glass surfaces show no significant change in the fraction of residues with the increase in humidity (remaining at ~55% for stainless steel and ~65% for uncoated glass). The final residue size formed on each surface shows a dependence on the humidity level as well as the initial droplet size for all surfaces (Figure S3). At a fixed humidity level, the residue size scales linearly with the initial droplet size with a slope varying from ~0.06 for uncoated glass to ~0.22 for stainless steel at 25% RH to ~0.08 for plastic and ~1.4 for stainless steel at 60% RH. The measurements on the stainless steel surface show the presence of two clusters that each scale differently with the initial droplet size at 60% RH. A cluster of large residues increasing at a higher rate as well as a smaller cluster that changes slowly with the initial droplet size. Note that we neglect the smaller size residues when estimating the linear trend line for stainless steel. We also observe a lower variation in the residue size at the higher humidity (within each type of residue for stainless steel). Finally, the smallest droplets that form residues decreases with increasing humidity (from 25% RH to 60% RH), albeit to different levels. The coated glass surface shows the highest variation from ~40 µm at 25% RH to ~5 µm at 60% RH, followed by the remaining three surfaces which show a drop of ~30 µm changing from ~40 µm to ~11 µm, ~12 µm and ~10 µm for the stainless steel, plastic and uncoated glass surfaces, respectively.

In contrast, as the humidity is reduced, the fraction of droplets that form residues decreases on all surfaces, with a maximum of 10% on coated glass at ~20% RH, with other surfaces indicating much smaller values. With a further drop in humidity to ~10% RH, none of the surfaces can form residues. The size of residues also indicates a strong dependence on the initial droplet size at each humidity investigated.

Interestingly, the steel surface at 60% RH shows two specific clusters corresponding to a larger and smaller residue types, each scaling differently with initial droplet size. In addition, all surfaces show a lower scatter in residue size at the higher humidity, possibly due to a reduction in formation of multiple residues, since surface tension makes it less likely for thicker films to breakup into pieces.

The water droplets are generated using distilled water with TSI 9302 nebulizer operated at an input pressure of 138 kPa which produces a 5.7 L/min output rate of droplets (mean diameter ~6.4 µm) which coagulate on the surface to produce a wide range of droplet sizes. Five different surface samples, including Fisher Scientific microscope glass slide, glass slide coated with RainX hydrophobic coating, plastic (3M polypropylene tape), copper (Hillman copper sheet) and 304 stainless steel samples, are selected for testing under an ambient temperature of 22 °C and humidity varying between 10% to 60% RH. The samples are placed with the test side facing up on an inverted microscope, connected with a Flare CMOS camera (2048 pixel × 1024 pixel sensor size) sampling at 30 frames/second. We used the nebulizer to generate droplets on the substrate (deposited size range 5 to 100 µm) and imaged them simultaneously under 10x magnification (1.21 mm × 0.64 mm field of view at 0.59 µm/pixel resolution) to capture the evaporation of liquid droplets and formation of the residues. The size of evaporating droplets at each time step and the corresponding residues are extracted from the 10x microscopic images manually using ImageJ, where the size is defined as the area-equivalent diameter. We conduct residue removability tests for each substrate through heating as well as wiping. For the former, we treat each surface with a heat gun (temperature of 60 °C at the surface) for 60 seconds and observe, both qualitatively and quantitatively, the change in the residue concentration. As for the latter, we wipe the surfaces with a Kimtech wipe for approximately 10 seconds, with minimal pressure. Finally, we test the long-term stability and durability of the residues on all surfaces (except copper) by capturing images at 10x magnification for 24 hours, at 1 hour increments, in an environment with relatively stable temperature (22 °C) and humidity (40% RH).

**Movie S1 (separate file).**

Droplet evaporating on coated glass forming a single residue

**Movie S2 (separate file).**

Droplet evaporating on coated glass with a film break up resulting in multiple residues.

**Movie S3 (separate file).**

Droplet evaporating on uncoated glass forming a single film type residue

**Movie S4 (separate file).**

Droplet evaporating on a coated glass exhibiting a film recoil near the end, decreasing the size of the residue

**Movie S5 (separate file).**

Droplet evaporating on a stainless steel forming an extended thin film type residue

**Movie S6 (separate file).**

Droplet evaporating on stainless steel that breaks up into multiple small residues